\documentclass[11pt,twoside]{article}

\usepackage{asp2006}
\usepackage{epsf}
\usepackage{psfig}
\usepackage{lscape}
\usepackage{graphicx}

\newcommand{\dlp}{\boldmath $\delta$\unboldmath{\bf P }}

\begin{document}
\title{Testing the potential and limitations of seismic data}   
\author{Orlagh L. Creevey}
\affil{Instituto de Astrof\'isica de Canarias, 
C/ V\'ia Lactea s/n, La Laguna 38205, Tenerife, Spain}

\begin{abstract} 
I describe the advantage of using singular value decomposition as a 
diagnostic tool for exploring the potential and limitations of 
seismic data.  
Using stellar models coupled with the expected errors in 
seismic and complementary data we can predict the precision in the 
stellar parameters.  
This in turn allows us to quantify if and to what extent we can 
distinguish between various descriptions  of the interior physical processes.  
This method can be applied to a wide range of astrophysical problems, 
and here I present one such example which shows that 
the convective core overshoot parameter can be constrained with one 
identified 
mode if the pulsating component is in an eclipsing binary system.

\end{abstract}



\section{Introduction}
It is of extreme benefit to have a reliable method to explore both 
the scientific potential {\it and} the limitations of a particular 
astrophysical system.
This paper is dedicated to discussing how to
 use some properties of singular value decomposition (SVD)
to do just this.
We can use 
SVD to do such things as 
predict the precision of global quantities, 
explore the impact of 
extending a data set from 3 months to 6 months (e.g. with Kepler), 
and probe the scientific benefit of complementary observables.
If we understand the potential of the system, then we can carefully choose
the scientific question that we would like to pursue.
As a specific example, we will test if an eclipsing binary system containing a 
pulsating component has sufficient information to allow us to constrain
the convective core overshoot parameter, $\alpha_{\rm ov}$.
This method of using SVD was exploited by \citet{bro94}, 
and more recently in the literature, by
\citet{mig05}, \citet{cre07} and \citet{cre08a}.
\citet{num92} while concentrating on SVD from a numerical point of view,
also briefly discusses some of these properties.

\section{Observational data and interpretation\label{sec:two}}
Normally in astrophysical problems, one is confronted with comparing 
a set of observational data with an astrophysical 
model for scientific interpretation.
This model should describe as best as possible the physical or chemical
processes observed.
The {\it input} to this model are a set of  
{\it parameters of the system}, which we shall denote by {\bf P}.  
Often we know what the parameters are, but
we do not know their quantities. 
For example, in a spectroscopic binary system, such parameters are the 
orbital period, the mass ratio and the inclination of the system with respect
to the observer.
Given a specific set of input {\bf P}, this model will produce a set 
of {\it expected observables}, which we denote by {\bf B}.  
In the binary system example,
{\bf B} could be the radial velocities of both components.
Scientific interpretation often then involves comparing 
the real observational data, {\bf O}, to {\bf B} to infer 
the values of {\bf P}.

If the system is linear, or if it can be described locally as linear,
the following equation can determine the {\it true} parameters
of the system, by calculating (iteratively) the parameter changes 
\dlp to make to an initial guess {\bf P$_0$}:
\begin{equation}
\boldmath\delta\unboldmath {\bf P = V}{\bf {W}^{-1}} {\bf U} \frac{{\bf O - B}}
{\boldmath\epsilon\unboldmath},
\label{eqn:dp}
\end{equation}
 where \boldmath$\epsilon$ \unboldmath are the 
measurement errors.
Here, we can see that the parameter changes needed to match the observations 
come from mainly a product of two terms:  the latter term in Eq.~\ref{eqn:dp}
is simply the 
scaled difference between the observations and the expected observables --- 
the {\it discrepancies} --- and the first term in Eq.~\ref{eqn:dp}
is the inverse of the 
sensitivity matrix (calculated from stellar models), expressed in its 
SVD form.
(This matrix is: 
\boldmath$\frac{\partial B}{\partial P}/\epsilon $ \unboldmath.)
The amount by which we need to change any parameter to successfully 
match {\bf O} with {\bf B} comes from a product
of the discrepancies in the observations, and a set of linear vectors 
(given by {\bf U} and {\bf V})
describing the relationship between each observable and each parameter.
The rest of this paper 
is not dedicated to solving these equations, but rather
to exploring the information contained in {\bf U} and 
{\bf V} to gain an understanding of the system under study.

\subsection{Relationship vectors versus the sensitivity matrix}
Let's concentrate on a more concrete example.  
Imagine that the system under study is a single isolated pulsating star.
The model consists of a stellar evolution and structure model, coupled
to an adiabatic oscillation code.  Here I use the Aarhus Stellar Evolution
Code 
\citep{jcd82,jcd07a}, 
coupled to the ADIabatic PuLSation 
code 
\citep{jcd07b}.
The {\bf P} are mass $M$, age $\tau$, initial mass fractions of the elements
$X$ and $Z$, 
and a mixing-length parameter to describe the outer convection zone $\alpha$.
The {\bf B} are the classical observables of effective temperature
$T_{\rm eff}$, luminosity $L_{\star}$ and 
metallicity $[M/H]$, and the average seismic quantities:
the {\it small} 
and {\it large frequency separations}, 
$\langle\delta\nu\rangle$ and $\langle\Delta\nu\rangle$. 
These quantities are the average values of the observed
$\delta\nu_{l,n} = \nu_{l,n} - \nu_{l+2,n-1}$ and
$\Delta\nu_{l,n} = \nu_{l,n} - \nu_{l,n-1}$ where $\nu_{l,n}$ are the 
oscillation modes of degree $l$ and radial order $n$.
The errors on each of the observations are typical: 200 K, 0.1 L$_{\odot}$, 
0.1, 1.3$\mu$Hz, 1.3$\mu$Hz.
\begin{table}
\begin{center}
\caption{Sensitivity matrix: 
partial derivatives of some stellar observables with respect to 
the global parameters, divided by the measurement errors\label{eqn:mat}}
\begin{tabular}{lcrrrrr}
\hline\hline
& &M & $\tau$& $X$ & $Z$ & $\alpha$\\
\hline
$T_{\rm eff}$ && 36.5 & 0.3 & -70.5 & -421.9 & 3.4\\
$L_{\star}$ && 58.5 & 50.5 & -71.1 & -390.0 & 6.2\\
$\log (Z/X)$ && 0.0 & 0.0 & -6.1 & 217.3 & 0.0\\
$\langle\Delta \nu \rangle$ && -3180.9 & -58.1 & 3913.82 & 18874.3 & 235.5\\
$\langle\delta\nu\rangle$ &   & -264.0 & -13.9 & 384.5 & 1628.7 & 4.18\\
\hline\hline
\end{tabular}
\end{center}
\end{table}

Table~\ref{eqn:mat} shows an example of the sensitivity matrix of the system
just described, taking as reference a star with parameters similar to the Sun.  
Each element is a partial derivative of one observable (specified in the 
leftmost column) with respect to one parameter (specified in the top row).
For example, the partial derivative of $L_{\star}$ with respect to $\tau$ 
is 50.5.
Note that these values take \boldmath$\epsilon$ \unboldmath 
into account, and will change if we assume some other values of 
\boldmath$\epsilon$\unboldmath.

If there were a discrepancy in only one of the observables, say, $\langle\delta\nu\rangle$,
 it 
is not so clear by reading this matrix which parameters and by how much 
each of these will need to change to reconcile {\bf O} with {\bf B},
mainly because $\langle\delta\nu\rangle$ is sensitive to each parameter.
However, representing this matrix in its SVD form allows us to do this
very simply. 

\begin{figure}
\center{\includegraphics[width=0.95\textwidth]{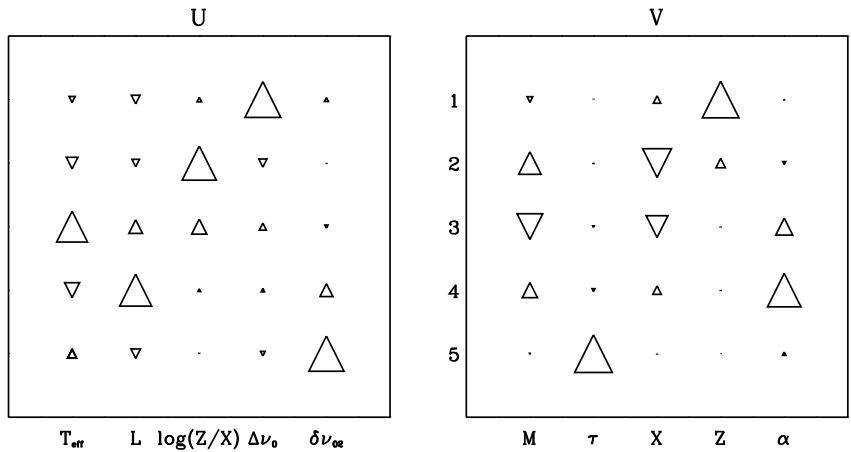}}
\caption{The decomposition matrices {\bf U} and {\bf V}, describing
the relationship between the stellar observables (left panel) 
and the stellar parameters (right panel) for a single 1 M$_{\odot}$ star.
\label{fig:uv}}
\end{figure}

Figure \ref{fig:uv}
is a graphical representation of the SVD of the matrix given 
in Table~\ref{eqn:mat}. 
In the decomposition both {\bf U} and {\bf V}
are orthonormal matrices that span, respectively, the observable and the 
parameter spaces.
Therefore 
each element is a value that varies between -1 and 1, and 
in Figure~\ref{fig:uv} each value is represented by a triangle, whose 
magnitude and direction is proportional to it.

Instead of thinking of these matrices as two separate matrices, we should
interpret them rather as sets of vectors that are related according to their
position in the matrix.
For example, row 1 in {\bf U} ({\it vector 1} or $U_1$) 
corresponds {\it only} to vector 1 in {\bf V}
and it is related by the first (highest) singular value $W_1$.

Now, we can investigate which of the parameters should be changed in order
to reconcile the discrepancy in the observable $\langle\delta\nu\rangle$.
Inspecting the column containing this observable on the left panel, 
we see that the largest component is contained in the 5th row of 
the matrix U ($U_5$).  
This implies that 
the parameter changes given by $V_5$ are those necessary
to reconcile the discrepancy in the observation.
However, $V_5$ (the bottom row on the right panel)
has mainly one non-zero value,
and this corresponds to $\tau$.
This means that in order to resolve the discrepancy in $\langle\delta\nu\rangle$, one would
need to adjust the value of $\tau$. 
The amount by which $\tau$ needs to be changed is proportional to the 
discrepancy in the observation 
and 
inversely proportional to the corresponding singular value, in this case
$1/W_5$.
$W_5$ is the smallest singular value, so $1/W_5$ is the largest inverse.
Now, we can also see that resolving discrepancies in the observables
that appear in vectors $U_1$, $U_2$, etc. 
would cause smaller \boldmath$\delta$\unboldmath{\bf P} 
than those in $U_5$ because $1/W_1$, $1/W_2$, etc. are much smaller
than $1/W_5$.   
This implies that the parameters that appear in the topmost 
vectors are only 
allowed to be adjusted by a small amount, while those in the lowest
vectors
can vary more, i.e.
the parameters appearing in $V_1$, $V_2$, etc.
have tighter constraints, and thus have the smallest uncertainties 
associated with them.  In fact, the uncertainties $\sigma_j$ for each 
$j$ parameter come in a neat and 
compact form:
\begin{equation}
\sigma_j^2 = \sum_{k=1}^N \frac{V_{jk}^2}{W_{kk}^2} .
\label{eqn:unc}
\end{equation}

Lets take as another example the observable $\langle\Delta\nu\rangle$. 
This observables 
appears mainly in $U_1$, implying that the 
parameter adjustments given by $V_1$ will resolve 
any discrepancy in this 
observation (if one exists).
In this case, we would need to 
increase $M$ and decrease $X$
by the same amount, while also increasing the value of $Z$ by a larger 
amount.
Again, the actual value by which we need to adjust these parameters is 
proportional to $1/W_1$ and the discrepancy in $\langle\Delta\nu\rangle$.
We can now begin to understand that these matrices show quite directly 
which observables contribute to determining each parameter.

As a third example, 
if the uncertainty in the mixing-length parameter $\alpha$
were quite large and it was in our interest to 
decrease it, given that 
$\alpha$ appears mainly in $V_3$ and $V_4$, a reduction in 
the errors on the observables appearing in $U_3$ and $U_4$ would 
cause the desired decrease in $\sigma(\alpha)$ --- 
these responsible observables are mainly 
$T_{\rm eff}$ and $L_{\star}$ (in this particular case).

One must also take some precaution with this interpretation: these results 
are sensitive to the observational errors that we assume, the set
of observables that we take into account, {\it and} the range of parameters
that we are studying.
Here we assume a single star with solar characteristics, but these 
relationships will change if the star is at a different 
evolutionary stage,  metallicity, and mass --- 
for a given reference set of parameters, 
these relationships extend to parameter ranges where we can still assume 
linearity with respect to the reference set. 

\subsection*{}
I showed briefly how we can interpret the decomposition matrices,
and how useful the properties of SVD are for understanding the 
system under study.
Not only can we investigate the role that each observable plays in 
determining the system parameters, but also SVD shows us if an observable
has no important role, i.e. if the observable is redundant (in the case of an 
over-determined system).
Being redundant implies that this observable can be used to 
test specific physical phenomena because they are independent of the stellar
model chosen to represent the star.

\begin{figure}
\center{\includegraphics[width=0.9\textwidth]{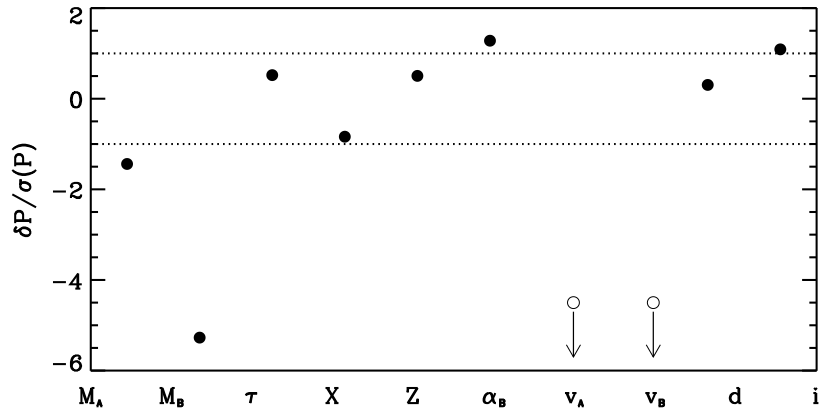}}
\caption{Parameter adjustments 
\boldmath$\delta$\unboldmath{\bf P}
needed to most adequately fit {\bf O} when
the model used for the inversion is incorrect.
The last four parameters are rotational velocity of component A, component B,
the distance to the system and the inclination. 
The values for $v_A$ and $v_B$ fall outside of the figure.\label{fig:dpar}}
\end{figure}  

\section{Determining the convective core overshoot parameter}
Let's suppose now that the system is a main sequence eclipsing spectroscopic
binary system with component masses of roughly 1.8 and 1.1 M$_{\odot}$ and
the 1.8  M$_{\odot}$ is a $\delta$ Scuti star.
If we consider both photometric and spectroscopic light curve observations, then
the observables will comprise of things such as the effective temperatures 
$T_{\rm eff\_A}$, 
and ratio T$_{\rm eff\_B}/T_{\rm eff\_A}$, 
inclination $i$, orbital period $\Pi$, radii $R_A$, $R_B$, mass estimates
$M_A \sin i$, $M_B \sin i$, and an identified oscillation mode $\nu$ from the 
$\delta$ Scuti component. 
An analysis of these observables following the discussion in 
Sec.~2 shows that all of the parameters of the 
system (now including two stars that we assume are coeval) are well-constrained
\citep{cre08b}.  
We pose the following question:
Can the single identified oscillation frequency be used to learn about 
the convective core overshoot parameter $\alpha_{\rm ov}$?

In order to set about answering 
this question, we simulated a set of observations {\bf O},  
from a model with  $\alpha_{\rm ov} = 0.3$.
We then used Eq.\ref{eqn:dp} iteratively to 
recover the input set 
of parameters, while using the correct model in the inversion. 
The parameters converged to the original set to 
within a small amount, we denote these by {\bf P$_{\rm F}$}.
We obtain the  uncertainties,  
\boldmath$\sigma$\unboldmath{\bf (P)} from
Eq.~\ref{eqn:unc}.

Now we change the inversion model to 
$\alpha_{\rm ov}$ = 0.2, and we try to recover the original 
observations using again, Eq.~\ref{eqn:unc}, 
while use the  
fit {\bf P$_{\rm F}$} as the initial guess.
Because we are using the incorrect physical model to fit the observations,
then 
we should expect to see some non-zero values of
\boldmath$\delta$\unboldmath{\bf P}. 
The question is, is the difference in the observables sufficient to be able
to detect?
We can argue that if there are \boldmath$\delta$\unboldmath{\bf P} 
that are larger than the 
\boldmath$\sigma$\unboldmath{\bf (P)}  calculated, then we conclude
that there is enough information in {\bf B} to detect an error in the 
physical assumptions.

Figure~\ref{fig:dpar} shows 
\boldmath$\delta$\unboldmath{\bf P}$/$\boldmath$\sigma$\unboldmath{\bf (P)}
for each {\bf P} after the inversion with the incorrect 
($\alpha = 0.2$) model.  The set of observables include one 
identified mode.
The horizontal dotted lines represent 
$\pm$1-\boldmath$\sigma$\unboldmath{\bf (P)}.  
It is clear that the eclipsing binary observables are capable of 
providing different 
{\bf P$_{\rm F}$} solutions 
based on diverging physical assumptions --- implying that there is 
sufficient information there to learn about $\alpha_{\rm ov}$.

\begin{figure}
\center{\includegraphics[width=0.9\textwidth]{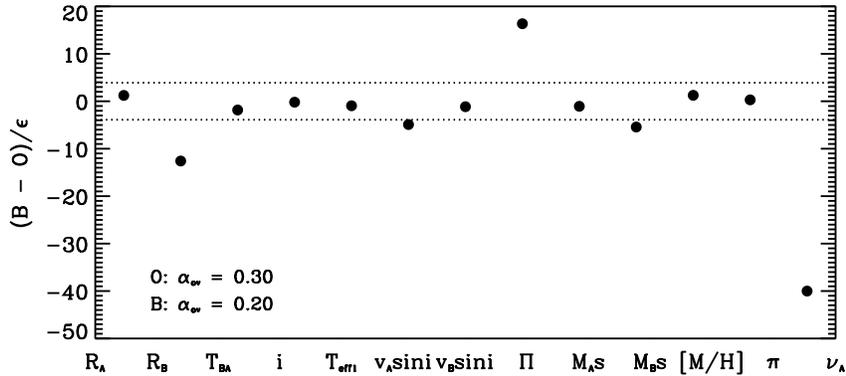}}
\caption{{\it Discrepancies} in {\bf O} when they are fit to 
the $\alpha_{\rm ov} = 0.2$ model.  The best set of {\bf P} result
in large deviations for some observables. \label{fig:dobs}}
\end{figure}

But how can we distinguish between the two fit parameter sets
given that both 
make physical sense?
Figure~\ref{fig:dobs} shows the new {\it discrepancies} in 
the observables, where {\bf B} are calculated from the $\alpha_{\rm ov} = 0.2 $
models (our assumption for the inversion) with the 
fit parameters corresponding to those in Fig.~\ref{fig:dpar}.
The large discrepancy in three of the observables indicates that our model
is incorrect.  
The identified mode is the observable that provides most evidence that 
$\alpha_{\rm ov} = 0.2$ is incorrect, its discrepancy is 40 times the 
allowed amount away from the observed value.
Repeating this for various values of $\alpha_{\rm ov}$ allows us to 
recover correctly the input value of 0.3.

\acknowledgements 
OLC is forever indebted to all of 
her PhD advisors: Timothy Brown,
Sebastian Jim\'enez-Reyes, Juan Antonio Belmonte, and Travis Metcalfe.
She is also grateful to Fernando P\'erez Hern\'andez 
for feedback on this manuscript.
Part of this work 
was conducted at the 
High Altitude Observatory, Boulder, Colorado.


\end{document}